\documentclass[prd,twocolumn,showpacs,groupedaddress,superscriptaddress,amsmath,amssymb]{revtex4-2} 
 
\newcommand{\vect}[1]{\boldsymbol{#1}}

\usepackage[utf8]{inputenc}
\usepackage[english]{babel}	

\usepackage{stfloats}

\usepackage{cmap}
\usepackage{bbm}

\usepackage{enumerate}
\usepackage{multirow}
\usepackage{float}
\usepackage{comment}

\usepackage{amsmath,amsfonts,amssymb,mathtools}	
\usepackage{euscript} 
\usepackage{mathrsfs}
\usepackage{mathtext}

\usepackage{color}			
\usepackage{graphicx}
\usepackage[export]{adjustbox}
\DeclareGraphicsExtensions{.png,.jpg}
\graphicspath{}	
\usepackage[section]{placeins} 

\usepackage{physics}
\usepackage{braket}
\usepackage{xparse}	

\usepackage{slashed}
\usepackage{indentfirst} 
\usepackage{enumitem}

\usepackage[colorlinks=true, filecolor=blue, citecolor=blue,urlcolor=black]{hyperref}
\usepackage{orcidlink}

\usepackage{url}
\usepackage{cleveref}
\date{\today}

\begin{document}

\title{Probing hidden sectors with a muon beam: implication of spin-0 dark matter mediators for muon $(g-2)$ anomaly and validity of the Weiszäcker-Williams approach}

\author{H.~Sieber\orcidlink{0000-0003-1476-4258}}
\email[\textbf{e-mail}:]{henri.hugo.sieber@cern.ch}
\thanks{Corresponding author}
\affiliation{ETH Zurich, Institute for Particle Physics and Astrophysics, CH-8093 Zurich, Switzerland}

\author{D.~V.~Kirpichnikov\orcidlink{0000-0002-7177-077X}}
\affiliation{Institute for Nuclear Research, 117312 Moscow, Russia}

\author{I.~V.~Voronchikhin\orcidlink{0000-0003-3037-636X}}
\affiliation{ Tomsk Polytechnic University, 634050 Tomsk, Russia}

\author{P.~Crivelli\orcidlink{0000-0001-5430-9394}}
\affiliation{ETH Zurich, Institute for Particle Physics and Astrophysics, CH-8093 Zurich, Switzerland}

\author{S.~N.~Gninenko\orcidlink{0000-0001-6495-7619}}
\affiliation{Institute for Nuclear Research, 117312 Moscow, Russia}

\author{M.~M.~Kirsanov\orcidlink{0000-0002-8879-6538}}
\affiliation{Institute for Nuclear Research, 117312 Moscow, Russia}

\author{N.~V.~Krasnikov\orcidlink{0000-0002-8717-6492}}
\affiliation{Institute for Nuclear Research, 117312 Moscow, Russia}
\affiliation{Joint Institute for Nuclear Research, 141980 Dubna, Russia}

\author{L.~Molina-Bueno\orcidlink{0000-0001-9720-9764}}
\affiliation{Instituto de Física Corpuscular, Universidad de Valencia and CSIC, Carrer del Catedrátic José Beltrán Martinez, 2, 46980 Paterna, Valencia, Spain}

\author{S.~K.~Sekatskii\orcidlink{0000-0003-4706-0652}}
\affiliation{Ecole Polytechnique Federale de Lausanne, CH-1015 Lausanne, Switzerland}

\begin{abstract}
In addition to vector ($V$) type new particles extensively discussed previously, both CP-even ($S$) and CP-odd ($P$) spin-0 Dark Matter (DM) mediators can couple to muons and be produced in 
the bremsstrahlung reaction $\mu^- + N \rightarrow \mu^- + N + S(P)$. Their possible 
subsequent invisible decay into a pair of Dirac DM particles, $S(P) \to \chi \overline{\chi}$, 
can be detected in fixed target experiments through missing 
energy  signature. In this paper, we focus on the case of experiments using high-energy muon beams. For this reason, we derive the  differential cross-sections involved using the phase space Weiszäcker-Williams approximation and compare them to the exact-tree-level calculations. The formalism derived can be applied in various experiments that could observe muon-spin-0 DM interactions. This can happen in present and future proton beam-dump experiments such as NA62, SHIP, HIKE, and SHADOWS; in muon fixed target experiments as NA64$\mu$, MUoNE and M3; in neutrino experiments using powerful proton beams such as DUNE. In particular, we focus on the NA64$\mu$ experiment case,  which uses a 160 GeV muon beam at the CERN Super Proton Synchrotron accelerator. We compute the derived cross-sections, the resulting signal yields and we discuss the experiment projected sensitivity to probe the relic DM parameter space and the $(g-2)_\mu$ anomaly favoured region considering $10^{11}$ and $10^{13}$ muons on target. 
\end{abstract}

\maketitle

\section{Introduction}
The Standard Model (SM) cannot explain the origin of dark matter (DM), although it makes up almost 
$\simeq 85\%$ of the Universe's matter~\cite{Planck:2018vyg}. The indirect evidence of DM are 
associated with the rotational velocities of galaxies, the cosmic structure of a large scale,  the 
anisotropy of the cosmic microwave background, and gravity 
lensing~\cite{Gelmini:2015zpa,Bergstrom:2012fi,Bertone:2004pz}. 
Nevertheless, the composition of DM continues to be one of the most challenging puzzles for  particle physics.  

Theoretically, well-motivated scenarios to explain the origin of Dark Matter as a thermal freeze-out relic involve the presence of feebly interacting light scalars from dark sectors (DS)~\cite{Chen:2018vkr,Berlin:2018bsc}. This framework addresses the origin of DM using a similar mechanism to the weakly interacting massive particles and could imply the existence of sub-GeV spin-0 DM mediators with feebly interaction strength~\cite{Agrawal:2021dbo}.

In addition, the observed low energy experimental anomalies such as the recently confirmed tension of $4.2\sigma$ \cite{Muong-2:2021ojo} in the measurement of the muon's anomalous magnetic moment~\cite{Aoyama:2020ynm}
\begin{equation}
\label{g-2Exp-Theor1}
     \Delta a_{\mu} \equiv a_{\mu}^{\rm exp} - a_{\mu}^{\rm th} =(251 \pm 59)\cdot 10^{-11},
\end{equation}
  has also motivated the existence of physics beyond the Standard Model and could be explained in DS framework~\cite{Muong-2:2006rrc}.  
We note that recent 
calculations~\cite{Borsanyi:2020mff,Ce:2022kxy,Blum:2023qou,Bazavov:2023has,ExtendedTwistedMass:2022jpw,RBC:2018dos} of 
the hadronic  vacuum polarisation contribution to $(g-2)_\mu$  shifts the anomaly to the level of 
$\Delta a_\mu = (183\pm 59)\cdot 10^{-11}$, that corresponds to a significance of $3.1\sigma$~\cite{WittigTalkMoriond} (for recent experimental results from CMD-3 collaboration see e.~g.~Ref.~\cite{CMD-3:2023alj}). 
In the present paper, we consider the result 
(\ref{g-2Exp-Theor1}) as a hint of new physics. In particular,
a possible solution to that discrepancy involves the introduction of a new weak coupling between the  standard matter and a light scalar DM mediator~\cite{Chen:2018vkr,Kahn:2018cqs,Chen:2017awl}. Other possibility to address the anomaly considers the case of a light vector mediator (see more details in \cite{Kirpichnikov:2021jev,Sieber:2021fu}).
This study focuses on the computation of the production cross sections of scalar ($S$) and pseudo-scalar ($P$) mediators after a high energy muon scatters off in a target (see e.~g.~Fig.~\ref{fig:Feynman}).
Our study is  particularly relevant to experiments involving high energy muon interactions with a fixed target such as muon experiments NA64$\mu$ at CERN \cite{Gninenko:2640930, Sieber:2021fu} or the proposal $M^3$ at Fermilab~\cite{Kahn:2018cqs}. Nevertheless, it can also be relevant for i) current and planned proton beam-dump experiments such as NA62 \cite{NA62:2022qes}, SHIP ~\cite{Rella:2022len}, HIKE~\cite{HIKE:2022qra}, SHADOWS~\cite{ShadowsLOI}, the ILC beam dump~\cite{Asai:2021xtg,Asai:2023dzs}, ii) muon beam-dump ~\cite{Cesarotti:2022ttv}, iii)  the MUonE experiment~\cite{GrillidiCortona:2022kbq} and iv) DUNE \cite{DUNE:2020ypp} aiming also to perform complementary searches to such hidden particles. In this manuscript, we take as an example the NA64$\mu$ experiment at CERN devoted to probe weakly coupled dark sectors with muons. 
 
 In the NA64$\mu$ experiment, a 160 GeV muon beam is directed to an electromagnetic calorimeter functioning as 
 an active target, where the spin-0 DM mediators are produced. The resulting particles carry away a 
 portion of the primary muon beam energy. The measurement of the primary muon missing momentum is the key feature of the experimental technique. 

In this paper, the production cross-section of spin-0 particles on the reaction $\mu N \to \mu N S(P)$ are derived. In particular, 
we show that the widely used Weizs{\"a}cker-Williams (WW) approach for the 
spin-0 production reproduces the exact tree level (ETL) cross sections with 
an accuracy at  the level of $\lesssim \mathcal{O} (5\%)$. Furthermore, a novel analytical formula for computing the differential cross-sections in the 
WW approximation has been obtained  in 
order to perform more accurate and less computationally demanding MC simulations of a dark boson emission. The results have been implemented in the Geant4-based Dark Matter simulation package DMG4 \cite{Bondi:2021nfp,Kirsanov:2023evm,KirsanovDMG4inPrepapration}. 
Additionally, we analyse the differential cross sections with respect to the recoil angles of the muon and spin-0 DM mediators relevant to obtain accurate and realistic signal yields in fixed target experiments.

 This paper is organised as follows:
 In section~\ref{SectMuonPhilicSetup}, we discuss the typical scenarios for spin-0 DM mediators.   
 In section \ref{ETLSection}, we calculate at ETL the total cross-section for spin-0 mediator production.
 In section \ref{WWCSSection}, we discuss the differential cross-sections for the angle
 and energy fraction of the outgoing particles in the WW approach. 
 In section~\ref{AnaluticalIntegralSection}, we derive novel analytical differential cross-sections for the emitted spin-0 
 mediator in WW approach.  In section~\ref{SectNumericalCS} we compare WW and ETL cross sections. 
Finally, in section~\ref{BoundsSection}, we evaluate the projected sensitivities for NA64$\mu$ experiment in leptophilic scenarios. We summarise our results and conclusions in section~\ref{SectionSummary}.


\section{A simplified muon-philic model
\label{SectMuonPhilicSetup}}
In this paper, we focus on lepton-specific spin-0 mediators that do not need to couple to neutrinos and assume
muon-specific couplings of (pseudo-)~scalar boson. The simplified
muon-philic spin-$0$ boson Lagrangians can be written for scalar,~$S$, and pseudo-scalar,~$P$, 
respectively as follows
\begin{align}
& \mathcal{L} \supset \mathcal{L}_{SM} + \frac{1}{2} (\partial_\mu S)^2 - \frac{1}{2} m_S^2 S^2 + 
g_S  S \overline{\mu} \mu, 
\label{ScalLagr1}
\\
& \mathcal{L} \supset \mathcal{L}_{SM} + \frac{1}{2} (\partial_\mu P)^2 - \frac{1}{2} m_P^2 P^2 
+ i   g_{P}    P \overline{\mu} \gamma_5 \mu, 
\label{PseudoScalLagr1}
\end{align}
where $\mathcal{L}_{SM}$ is the SM Lagrangian, $g_{S(P)}$ is the coupling strength to muons and $m_{S(P)}$ the mass of the mediator. The extension to the Dark Sector can be 
introduced through the benchmark couplings to Dirac DM fermions 
\begin{align}
\label{ScalarToDMcoupl}
& \mathcal{L} \supset  \overline{\chi}( i \gamma^\mu \partial_\mu - m_\chi  )\chi + g_S^\chi S \, \overline{\chi}\chi ,  
\\
& \mathcal{L} \supset    \overline{\chi}( i \gamma^\mu \partial_\mu - m_\chi  )\chi + i g_P^\chi P \, \overline{\chi} \gamma_5 \chi ,
\label{PseudoScalarToDMcoupl}
\end{align}
where $m_\chi$ is a mass of DM particle, $g_S^\chi$ and $g_P^\chi$ are the typical DM couplings to scalar and pseudo-scalar mediators 
respectively. Moreover, we assume that the invisible decay of $S(P)\to \chi \overline{\chi}$ will be the dominant channel. This means that we focus only on the benchmark regime $m_{S (P)}\gtrsim  2 m_\chi$ and 
$g_{S(P)}^{\chi} \gg g_{S (P)}$ in the present study. 

We also note that scalar couplings~(\ref{ScalLagr1}) can be originated from flavour specific Lagrangian of higher 
dimensions in Higgs extended sectors~\cite{Batell:2016ove} that can be probed by accelerator-based 
experiments~\cite{Chen:2018vkr,Kahn:2018cqs,Chen:2017awl}.  For pseudo-scalar benchmark couplings (\ref{PseudoScalLagr1}) we address the reader to Ref.~\cite{Cheung:2022umw}, where muon-specific ALPs signatures were studied in detail in the light of atmospheric probes of ALPs  using Cerenkov detectors near the Earth's surface. 

The one-loop leading order contributions from scalars to the $(g-2)_\mu$  are obtained through the Yukawa-like interaction and are given by  \cite{Chen:2015vqy,Leveille:1977rc,Lindner:2016bgg,Kirpichnikov:2020tcf}
\begin{align}
&    \label{eq:delta_a_S}
    \Delta a_S=\frac{g_S^2}{8\pi^2}\int_{0}^{1}dx\ \frac{m_\mu^2(1-x)(1-x^2)}{m_\mu^2(1-x)^2+m_S^2x},
\end{align}
where we defined $g_S=e\epsilon_S$, with $e=\sqrt{4\pi\alpha}$ the electric charge and $\alpha\simeq 1/137$
the fine-structure  constant. In the case where $m_S/m_\mu\rightarrow0$, $g_S=(3.63\pm0.43)\times10^{-4}$.

On the other side, the one-loop contribution of the muon-philic pseudo-
scalar boson to $\Delta a_\mu$ is negative, so the CP-odd 
spin-0 mediator can not accommodate the explanation of the $(g-2)_\mu$ anomaly~\cite{Cheung:2022umw}. 
For completeness, we refer the reader to \cite{Kirpichnikov:2021jev} for a 
discussion of the contribution of the vector-boson. In addition, for recent progress on probing leptophilic dark 
sector see also Refs.~\cite{Rella:2022len,Moroi:2022qwz,Forbes:2022bvo,Balkin:2021rvh}.

\begin{minipage}{\linewidth}    
\begin{figure}[H]
    \centering
    \includegraphics[width=0.95\textwidth]{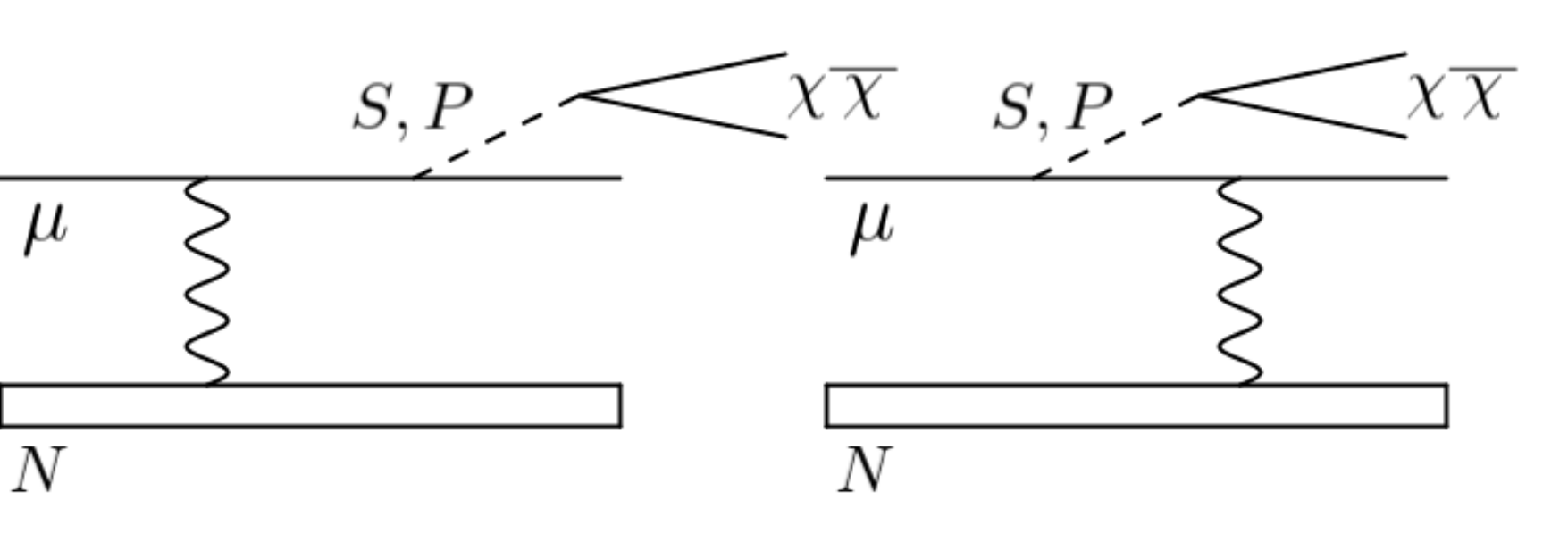}
    \caption{\label{fig:Feynman} Diagrams describing spin-0 DM mediator production via bremsstrahlung $\mu N \to \mu N S(P)$, followed by invisible decay $S(P)\to \chi \overline{\chi}$. 
    }
\end{figure}
\end{minipage}

\section{The exact tree-level calculation
\label{ETLSection}}
In the following, we discuss the computations of the exact-tree-level production cross-sections for both, a light 
scalar and pseudo-scalar muon-philic boson. We follow the notations of \cite{Liu:2016mqv,Gninenko:2017yus}. We 
refer to the kinematic variables of the
process $\mu^{-}(p)+N(P_i)\rightarrow\mu^{-}(p')+N(P_f)+H(k)$ from our previous 
work~\cite{Kirpichnikov:2021jev}. Here we denote via $H=(S,P)$ the general muon-specific CP-even and CP-odd spin-0 boson.
Let us recall the definition of the double-differential cross-section \cite{Liu:2016mqv}
\begin{align}
    \label{eq:etl_d2s}
  \frac{d\sigma}{dx\, d\!\cos\theta_{H}} \Bigr|_{ETL}\!\!\!
  \!\!\!\!=\!\frac{\!\epsilon_H^2\alpha^3|\mathbf{k}|\!E_\mu}{|\mathbf{p}||\mathbf{k}\!-\!\mathbf{p}|}
  \!\!\!\! \int\limits_{t_\text{min}}^{t_\text{max}}\!\!\!\frac{dt}{t^2}G_2^{el}(t)  \!\! \! 
  \int\limits_{0}^{2\pi}\!\!\frac{d\phi_q}{2\pi}\frac{|\mathcal{A}_{2\rightarrow 3}^{H}|^2}{8M^2}, 
\end{align}
where $t_\text{min}$ and $t_\text{max}$  the minimum and maximum momenta transfer and $G_2^{el}(t)$ the squared
elastic form factor as defined in \cite{Liu:2016mqv,Liu:2017htz,Kirpichnikov:2021jev}, $\phi_q$ is the axial 
angle of the three momentum transfer to the nucleus $ \vect{q} = \vect{P}_i -\vect{P}_f$ defined in the polar frame in Ref.~\cite{Liu:2016mqv}. The amplitude squared associated to the production of 
a (pseudo-)~scalar boson is calculated using the FeynCalc package \cite{Mertig:1990an} embedded in the Wolfram-
language-based Mathematica package \cite{Mathematica}. With similar kinematics as defined for the vector $V-$boson, 
we obtain in the case $H=S$
    \begin{align}
    \label{AS2}
  &      |\mathcal{A}_{2\rightarrow 3}^{S}|^2=\frac{1}{\tilde{u}^2\tilde{s}^2}\big\{4(4m_\mu^2-m_S^2)(\mathcal{P}\cdot p')^2\tilde{s}^2
   \\
        &-\! 4[t(\mathcal{P}\cdot p)^2\! -\! 2(4m_\mu^2\! -\! m_S^2\! +\! t)(\mathcal{P}\! \cdot\!  p)(\mathcal{P}\! \cdot\!  p')\! +\! t(\mathcal{P}\!\cdot\!  p')^2]\tilde{s}\tilde{u} \nonumber
        \\
        &+\!4(4m_\mu^2\!-\!m_S^2\!)(\mathcal{P}\!\cdot\! p)^2\tilde{u}^2\!+\!\mathcal{P}^2[\tilde{s}\!+\! \tilde{u}]^2[(m_S^2\!-\!4m_\mu^2)t\!+\!\tilde{s}\tilde{u}]\big\}, \nonumber
    \end{align}
as well as for the pseudo-scalar particle, $H=P$,
    \begin{align}
    \label{AP2}
        & |\mathcal{A}_{2\rightarrow 3}^{P}|^2=\frac{1}{\tilde{u}^2\tilde{s}^2}\big\{8(-m_P^2+t)(\mathcal{P}\cdot p)(\mathcal{P}\cdot p')\tilde{s}\tilde{u}
        \\
       & -4(\mathcal{P}\cdot p)^2\tilde{u}(t\tilde{s}+m_P^2\tilde{u}) \nonumber
        \\
        &-4(\mathcal{P}\cdot p')^2\tilde{s}(t\tilde{u}+m_P^2\tilde{s})+\mathcal{P}^2(\tilde{s}+\tilde{u})^2(m_P^2t+\tilde{s}\tilde{u})\big\}, \nonumber
    \end{align}
for which the relevant Mandelstam variables and dot products read
    \begin{equation}
        \begin{split}
        \tilde{s}&=(p'+k)^2-m_\mu^2=2(p'\cdot k)+m_H^2\,,
        \\
        \tilde{u}&=(p-k)^2-m_\mu^2=2(p\cdot k)+m_H^2\,,
        \end{split}
    \end{equation}
    \begin{align}
  &      \mathcal{P}^2\!=\!4M^2\!+\!t(p'\!+\!k)^2\,,
        \;
        (\mathcal{P}\!\cdot\! p)=2ME_\mu\!-\!(\tilde{s}\!+\!t)/2\,,
       \\
   &     \mathcal{P}\cdot p'=2M(E_\mu-E_{H})+(\tilde{u}-t)/2\,,
    \end{align}
with $\mathcal{P}_\mu=(P_i+P_f)_\mu$, $P_i=(M,0)$ being the nucleus four-momentum in the laboratory frame, 
$P_f=(P_f^{0},\mathbf{P}_f)$ is its outgoing momentum. The resulting squared matrix elements 
Eqs.~(\ref{AS2}) and~(\ref{AP2}) coincide with those given in Refs.~\cite{Liu:2016mqv} and~\cite{Liu:2017htz}, 
implying replacement of the electron with muon, i.e.  $m_e \to m_\mu$.

\section{The WW approximations for the (pseudo-)scalar emission cross-sections
\label{WWCSSection}}
In this section, we use the Weizs{\"a}cker-Williams (WW) approximation to compute the 
double-differential production cross-sections for $H$, assuming that the energy of the 
incoming muon is much larger than both $m_\mu$ and $m_H$. In this approach, the flux of virtual photons 
from the moving charged particles can be treated as a plane wave and approximated by real photons. 

We follow the same procedure as the one described in \cite{Kirpichnikov:2021jev}. In particular, for the choice of 
$(x, \theta_H)$ and $(y,\psi_\mu)$ variables, the WW-approximated quantities read respectively
\begin{align}
\label{eq:WWxtheta}
& \frac{d \sigma^H_{2\to 3}}{d x d \cos\theta_H} \Bigr|_{WW}\! \simeq\! \frac{\alpha \chi}{ \pi (\!1\!-\!x\!)} E_\mu^2 x \beta_H  \frac{d \sigma^H_{2\to 2} }{d (pk)} \Bigr|_{t=t_\text{min}}, 
\\
& \frac{d \sigma^H_{2\to 3}}{d y d \cos \psi_\mu} \Bigr|_{WW} \simeq \frac{\alpha \chi }{\pi (\!1\!-\!y\!)} E_\mu^2 y \beta_{\mu'} \frac{d \sigma^H_{2\to 2} }{d (pk)} \Bigr|_{t=t_\text{min}}, 
\label{eq:WWypsi}
\end{align}
where $\beta_{H} = (1- m_H^2/(x E_\mu)^2)^{1/2}$ and $x=E_H/E_\mu$ are the typical velocity of the produced hidden boson and  
its energy fraction respectively,  $\beta_{\mu} = (1- m_\mu^2/(y E_\mu)^2)^{1/2}$ and $y=E_{\mu'}/E_\mu$ are the typical velocity of the recoil muon and its energy fraction respectively, $\psi_\mu$ and $\theta_H$ are the recoil angles of outgoing muon and the
$H$-boson respectively. The expression of the photon flux $\chi$ is given by (\ref{ChiViaC}) in Sec.~\ref{AnaluticalIntegralSection} below. The cross-section of the process $\mu \gamma \to \mu H$ has the following form
\begin{equation}
\frac{d \sigma_{2\to 2}^{H} }{ d (pk)} = \epsilon^2_H \alpha^2 \frac{2 \pi }{\tilde{s}^2}   
\left| \mathcal{A}^H_{2\to 2 } \right|^2,
\label{dsdpkGeneralH1}
\end{equation}
where the squared amplitudes read
\begin{align}
\left| \mathcal{A}^S_{2\to 2} \right|^2\!\!\! =\!2(m_S^2 \!- \!4 m_\mu^2)\!\! \left[ \!\!\left(\!\frac{\tilde{s}\!+\!\tilde{u}}{\tilde{s} \tilde{u}}\!\right)^2
\!\!\!m_\mu^2\!
- \!\frac{t_2}{\tilde{s} \tilde{u}}\!\right]\!\! -\! \frac{(\tilde{s}\!+\!\tilde{u})^2}{\tilde{s} \tilde{u}}
\!,
\label{A2to2S} 
\\
 \left| \mathcal{A}^P_{2\to 2} \right|^2 = 
2 m_P^2 \left[ \left(\frac{\tilde{s}+\tilde{u}}{\tilde{s} \tilde{u}}\right)^2 m_\mu^2
\!- \! \frac{t_2}{\tilde{s} \tilde{u}}\! \right]\! -\! \frac{(\tilde{s}+\tilde{u})^2}{\tilde{s} \tilde{u}}.
\label{A2to2P} 
\end{align}
We note that for the $(x,\theta_H)$ - plane
one has the following expressions for the Mandelstam variables
\begin{equation}
\tilde{s} \simeq U/(1-x), \quad U = E_\mu^2 \theta_H^2 x + m_H^2 (1-x)/x +m_\mu^2 x,  
\end{equation}
\begin{equation}
 \tilde{u} \simeq - U, \quad
t_2 = -x U/(1-x) +m_H^2.     
\end{equation}
On the other hand, for the $(y,\psi_\mu)$ - plane the Mandelstam variables read 
\begin{equation}
\tilde{s}\simeq  \tilde{t}/(1-y), \quad \tilde{u} \simeq - y \tilde{t}/(1-y), \quad 
\tilde{t} \simeq  m_H^2 - t_2, 
\end{equation}
\begin{equation}
    t_2 \simeq  -[E_\mu^2 \psi_\mu^2 y + m_\mu^2 (1-y)/y +m_\mu^2 y ]+ m_\mu^2. 
\end{equation}
 Let us also remark on the typical energy fractions  of the outgoing muon and $H$ boson in the process $\mu N \to \mu N H$ for certain benchmark kinematics. 
 The lowest possible energy of the produced $H$-boson implies that $x_\text{min} \simeq 
 m_{H}/E_\mu \lesssim x$, i.~e. in this 
 case the spin-0 particle is produced with zero three-momentum,  $|\mathbf{k}|=0$. 
 This means also that almost all energy 
 of the initial muon is transferred to the outgoing muon, which leads to the typical bound 
 $y \lesssim y_\text{max} \simeq 1- m_{H}/E_\mu$. 
 On the other hand, if the initial muon transfers its maximal energy to spin-0 boson, then we get 
 $y\gtrsim y_\text{min} \simeq m_{\mu}/E_\mu$ and $x \lesssim  x_\text{max}\simeq 1- m_\mu/E_\mu$. 
 
\section{Analytical integration of the WW approximation over the angle $\theta_H$
 \label{AnaluticalIntegralSection}}
In the WW approach, the lower bound of the flux integral $t_\text{min}$ depends on both the fractional energy $x$ and 
the emitted angle $\theta_H$ of the boson mediator. Although WW provides more accurate results than its improved approach (IWW), the integration of the double-differential cross-section is still computationally expensive, to sample a sufficiently large number of MC events~\cite{Bondi:2021nfp}. In this work, we perform an explicit integration 
over $\theta_H$ to obtain an analytical expression for $d\sigma^H_{2\rightarrow3}/dx$. We emphasize that this result can 
also be expanded to the light $V$ vector boson case. 

    The formula for the differential cross-section can be rewritten in the following form
	\begin{align}
 \label{dsdx22}
	&	 \frac{d\sigma^H_{2\to3}}{dx}\Bigr|_{WW} 
	\!\!\!=\!
		\epsilon_H^2 \alpha^3 
		\sqrt{x^2\! - \!\frac{m_{H}^2}{E_{\mu}^2}}\!\!
		\frac{1\! - \!x}{x}\!\!\!
	\int\limits_{u_\text{min}}^{u_\text{max}}\!\!\! du
			\frac{\left| \mathcal{A}_{2 \rightarrow 2}^{H}\right|^2\!\! \chi}{u^2},
	\end{align}
	where the limits of integration over the Mandelstam variable are:
	\begin{align}	
	&	u_\text{max} = - m_{H}^{2}(1 - x)/x - m_{\mu}^{2}x,
	\\
	 &	u_\text{min} = - x E_{\mu}^{2} (\theta^\text{max}_H)^2 - m_{H}^{2}(1 - x)/x - m_{\mu}^{2}x,
	\end{align}
	where $\theta_H^\text{max}$ is the typical maximal angle between the initial muon and the emission
	momentum of the $H$ boson. Numerical analysis show (see e.~g.~Sec.~\ref{SectNumericalCS} below) that
	for the ultra-relativistic muons expected at NA64$\mu$ one can set $\theta_H^\text{max} \simeq 0.1$.  
	It is worth noticing that in Eq.~(\ref{dsdx22}) we imply  $d \cos\theta_H \simeq du/(2 x E_\mu^2)$ in order to introduce a new  variable of the integration $u$ instead of $\cos \theta_H$. The  transition amplitude squared then reads
	\begin{equation}
 \label{AViaC}
		\left| \mathcal{A}_{2 \rightarrow 2}^{H}\right|^2(x, u)
	=
        C^{H}_{1} 
    +   C^{H}_{2}  \frac{1}{u}
    +   C^{H}_{3}  \frac{1}{u^2},
	\end{equation}
	where the coefficients $C_i^H$ are
\begin{align}
 &	C^{S}_{1}  =  C^{P}_{1}  =  \frac{x^2}{1 - x}, \, C^{S}_{2}  = 2 \left( m_S^2 - 4 m_{\mu}^2\right) x,
  \\
&   	 C^{P}_{2}   =	 2 m_P^2 x, \, C^{P}_{3}   = 2 m_P^2	\left( m_P^2 (1 - x) + m_{\mu}^2 x^2 \right),	
  \\
& C^{S}_{3}   = 2 \left( m_S^2 - 4 m_{\mu}^2\right) \left( m_S^2 (1 - x) + m_{\mu}^2 x^2 \right).
\end{align}
For completeness, we also derive the coefficients for the vector boson  emission. 
In particular, for the case of $H=V$  these quantities  read explicitly in the following form 
\begin{align}
& 		C^{V}_{1}  
	= 
		2 \frac{(2 - 2 x + x^2)}{1 - x}, \quad
		C^{V}_{2}  
	=	
		4 \left( m_{V}^2 + 2 m_{\mu}^2\right) x,
\\
& 		C^{V}_{3}  
	=
		4 \left( m_{V}^2 + 2 m_{\mu}^2\right)
		\left( m_{V}^2 (1 - x) + m_{\mu}^2 x^2 \right).
\end{align}
	The flux of virtual photons $\chi$ in the Weizsacker-Williams approximation can be expressed via the typical elastic atomic form-factor in the following form
	\begin{multline}
	    \chi
	= 
	    Z^2 \! \! \int\limits_{t_\text{min}}^{t_\text{max}} \frac{t - t_\text{min}}{t^2} \left( \frac{t}{t_a + t} \right)^2 \!\!  \left( \frac{t_d}{t_d + t} \right)^2\!\! dt
	= 
	    C^{\chi}_{1} 
	\\ +
	    C^{\chi}_{2}  u^2
\!	+ \!  C^{\chi}_{3} \!  \ln{\left(\! \frac{u^2 g^2 \! + \!t_d}{u^2 g^2 \! + \!t_a}\!\right) }
	\!+ \!  C^{\chi}_{4}  u^2 
	    \ln{\! \left(\!  \frac{u^2 g^2  \!+\! t_d}{u^2 g^2 \! + \!t_a}\! \right) },
     \label{ChiViaC}
	\end{multline}
	where $Z=82$ is the atomic number of the lead target of NA64$\mu$, $\sqrt{t_{a}} = 1/R_a$ is a momentum transfer associated with nucleus 
 Coulomb field screening due to the atomic
electrons, with $R_a$ being a  typical  magnitude of the atomic radius $R_a = 111 Z^{-1/3}/m_e$, 
$\sqrt{t_{d}} = 1/R_n$ is the typical momentum associated  
with nuclear radius $R_n$, such that $R_n\simeq 1/\sqrt{d}$ and $d = 0.164 A^{-2/3}  \text{GeV}^2$,
$A=207$ is the atomic mass number of the lead target,
$t_\text{min} = g^2  u^2$ is minimal transfer momentum, here we denote 
 $g  = 1/(2 E_{\mu} (1 - x))$ for simplicity.
   Typically the maximal transfer momentum $t_\text{max}$ 
  is chosen to be $ t_\text{max}=m_\mu^2+m_H^2$ in~\cite{Liu:2016mqv,Liu:2017htz}, 
 however, the numerical calculations
  reveal that $t_\text{max}$ can be set as large as $t_\text{max} \gtrsim t_d$ in order to achieve a 
  better  accuracy for WW approach. The coefficients $C_1^\chi, C_2^\chi, C_a^\chi$ and $C_4^\chi$ 
  in~(\ref{ChiViaC}) are collected in   Appendix~\ref{AppSect}.

By	substituting Eqs.~(\ref{ChiViaC}) and (\ref{AViaC}) into the differential cross section (\ref{dsdx22}) one can 
obtain the following expression:
	\begin{multline}
		\left( \frac{d\sigma}{dx}\right)_{WW}\!\!\!\!\!
	=\!
		\epsilon_H^2 \alpha^3\!\! 
		\sqrt{x^2\! - \!\frac{m_{H}^2}{E_{\mu}^2}}
		\frac{1\! -\! x}{x} \!\!\!
        \int\limits_{u_\text{min}}^{u_\text{max}} \!\!\!
		\left\lbrace  
			C^{H}_{1} C^{\chi}_{2}  
		\!+\!	\frac{C^{H}_{2} C^{\chi}_{2} }{u}
		\right. \\
 +
			\frac{	C^{H}_{1} C^{\chi}_{1}  
				 + 	C^{H}_{3} C^{\chi}_{2} }{u^2}
		+	\frac{C^{H}_{2} C^{\chi}_{1} }{u^3}
		+	\frac{C^{H}_{3} C^{\chi}_{1} }{u^4}
		+ \\ + \left.		
			\left[
			C^{H}_{1} C^{\chi}_{4}    
		+	\frac{C^{H}_{2} C^{\chi}_{4} }{u}
		+	\frac{	C^{H}_{1} C^{\chi}_{3}  
			     +	C^{H}_{3} C^{\chi}_{4} }{u^2}
		+ \right. \right. \\ \left. \left.
			\frac{C^{H}_{2} C^{\chi}_{3} }{u^3}
		+	\frac{C^{H}_{3} C^{\chi}_{3} }{u^4}
			\right]
			\ln{\left( \frac{u^2 g^2  + t_d}{u^2 g^2  + t_a}\right) }
		\right\rbrace   
		du. \label{ResCsViaI}
	\end{multline}
	The differential cross section with respect to $x$,  can be represented as the 
 the sum of six typical  terms
	\begin{align}
        \label{eq:analytical_dsdx_WW}
	\left( \!\frac{d\sigma}{dx}\!\right)_{WW}\!\! \!\!\!\!
	=
		\!\epsilon_H^2 \alpha^3 \!\!
		\sqrt{\!x^2\! -\! \frac{m_{H}^2}{E_{\mu}^2}}
 \frac{1 \!-\! x}{x}
		\sum\limits_{i = 1}^{6} \Delta I^H_{i}(x, u),
	\end{align}
where $\Delta I^H_{i}(x, u)  = I^H_i(x,u_\text{max})-I^H_i(x,u_\text{min})$. The functions $I^H_i(x,u)$ are described in Appendix~\ref{AppSect}. Alternatively, one can also exploit the state-of-the-art MadGraph~\cite{Alwall:2014hca} or CalcHEP~\cite{Belyaev:2012qa} packages with appropriate atomic form-factor implementation~\cite{Chen:2017awl,Marsicano:2018vin,Forbes:2022bvo,Zhevlakov:2022vio,Eichlersmith:2022bit,Arefyeva:2022iba,Ariga:2023fjg}. That analysis, however, is beyond the scope of the present paper.




\begin{widetext}
\begin{minipage}{\linewidth}
\begin{figure}[H]
\centering
\includegraphics[width=0.8\textwidth]{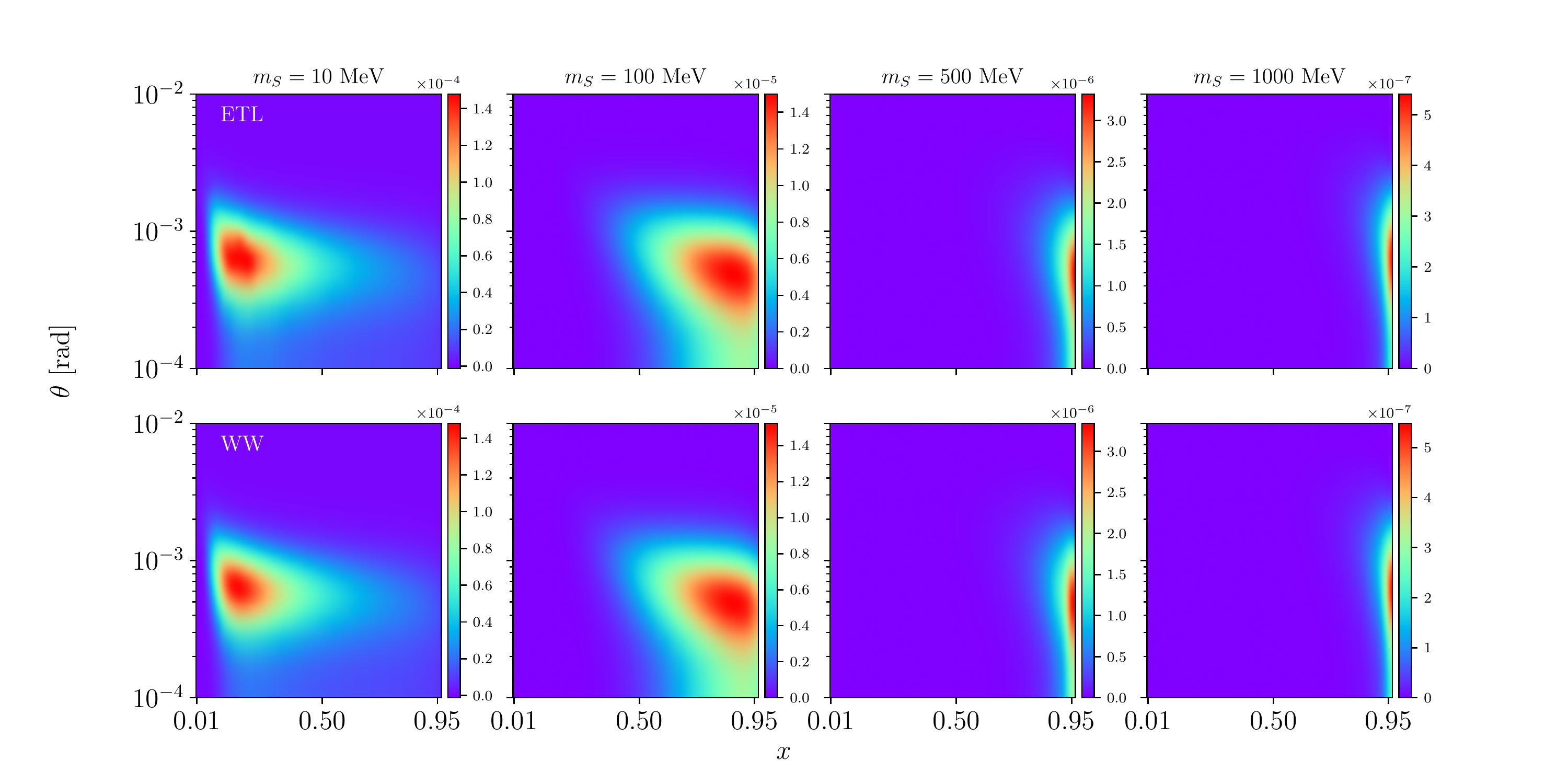}
\caption{\label{fig:comparison_ETL_WW_scalar_d2s}Top: double-differential cross-section at ETL in the $(x,\theta_S)$ space. The expression of Eq. \eqref{eq:etl_d2s} is integrated using both the angular parametrisation from \cite{Davoudiasl:2021mjy} and MC integration \cite{galassi2018scientific}. Bottom: double-differential cross-section in the WW approach for the $(x,\theta_S)$ variables (see Eq. \eqref{eq:WWxtheta} with an amplitude squared as defined in Eq. \eqref{A2to2S}). The mass range spans from 10 MeV to 1 GeV. The mixing strength is $\epsilon_S=10^{-4}$.}
\end{figure}

\begin{figure}[H]
\centering
\includegraphics[width=0.8\textwidth]{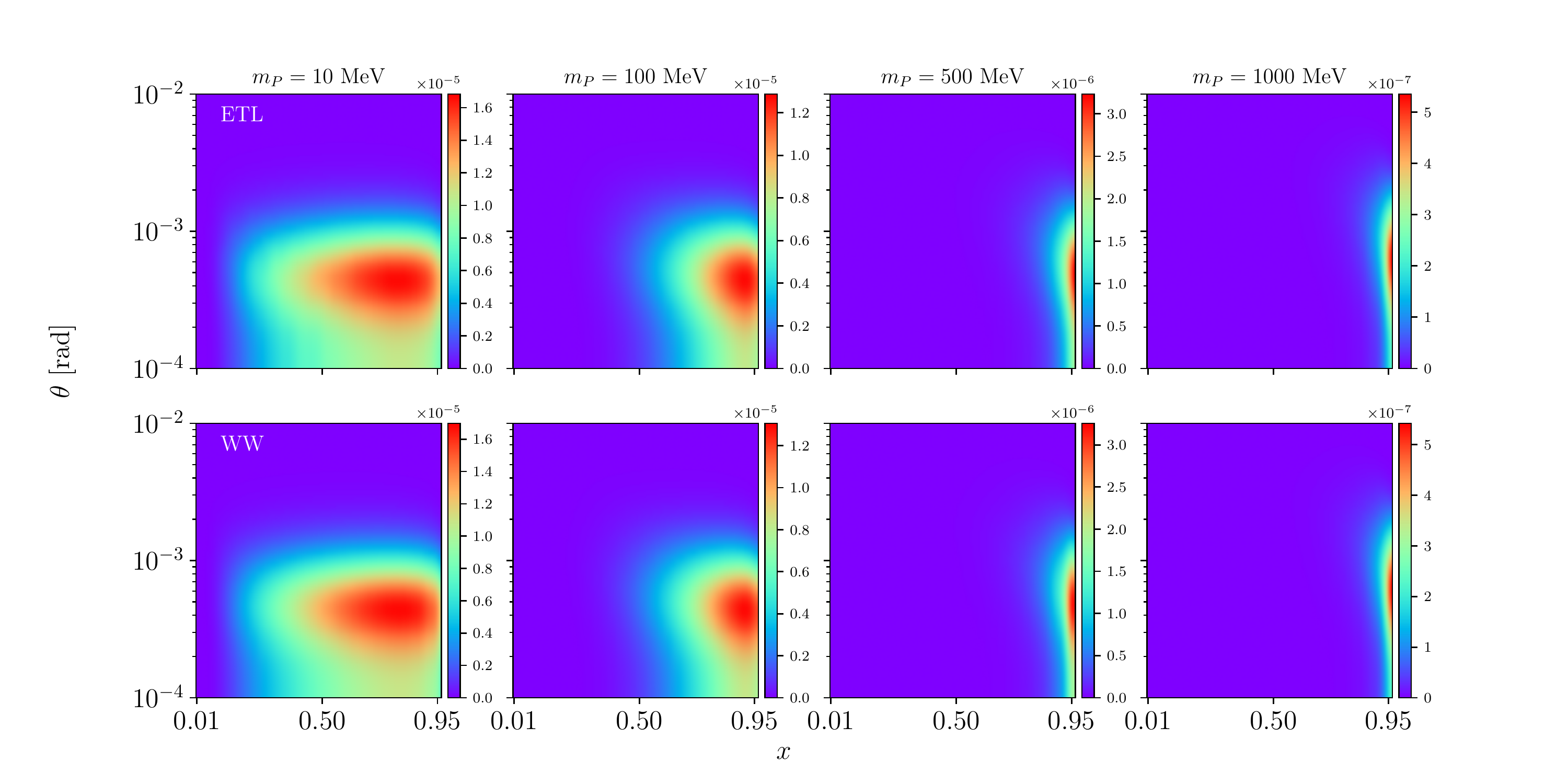}
\caption{\label{fig:comparison_ETL_WW_pseudoscalar_d2s} The same as in Fig.~\ref{fig:comparison_ETL_WW_scalar_d2s} but for pseudo-scalar field $P$. }
\end{figure}
\end{minipage}
\end{widetext}

\section{Numerical integration of the cross-sections
\label{SectNumericalCS}} 

The comparison between the expression for $d\sigma/dxd\cos\theta_H$ at ETL and in the WW approximation is 
shown for both scalar and pseudo-scalar mediators in Figs.~\ref{fig:comparison_ETL_WW_scalar_d2s} 
and \ref{fig:comparison_ETL_WW_pseudoscalar_d2s}.  The $H$ mass spans from 10 MeV to 1 GeV. Following the nominal beam energy of the NA64$\mu$ experiment, the initial state muon energy is set to $E_\mu=160\,~\mbox{GeV}$. The integration over $(t,\phi_q)$ in Eq.~\eqref{eq:etl_d2s} is performed using the parametrisation of \cite{Davoudiasl:2021mjy} to integrate out $\phi_q$. The integral over $t$ is computed numerically through Monte Carlo integration \cite{galassi2018scientific}. Only the region of phase space $(x,\theta_H)$ where the double-differential cross-section contributes more is shown. Both the complete calculation (ETL) and the WW results have the double differential cross-section peak at the same order of magnitude. Additionally, from both Fig. 
\ref{fig:comparison_ETL_WW_scalar_d2s} and \ref{fig:comparison_ETL_WW_pseudoscalar_d2s} it can be seen that $\theta_H$ is 
constant around $\sim5\times10^{-4}$ as expected from the typical emission angle $\sim m_\mu/E_\mu$, which is independent of $m_H$. 
To perform the comparison between the ETL and WW approximated results, Eq. \eqref{eq:etl_d2s} is integrated over $\theta$ 
and compared to the expression in Eq. \eqref{eq:analytical_dsdx_WW}. The results in the case $H=S$ and $H=P$ 
are shown respectively 
in Fig.~\ref{fig:comparison_ETL_WW_scalar_ds} and 
Fig.~\ref{fig:comparison_ETL_WW_pseudoscalar_ds}. In both cases the integrated differential cross-section relative 
error with respect to the exact calculation is below $\lesssim\mathcal{O}(5\%)$ for the full mass range. 

\begin{widetext}
\begin{minipage}{\linewidth}
\begin{figure}[H]
\centering
\includegraphics[width=0.8\textwidth]{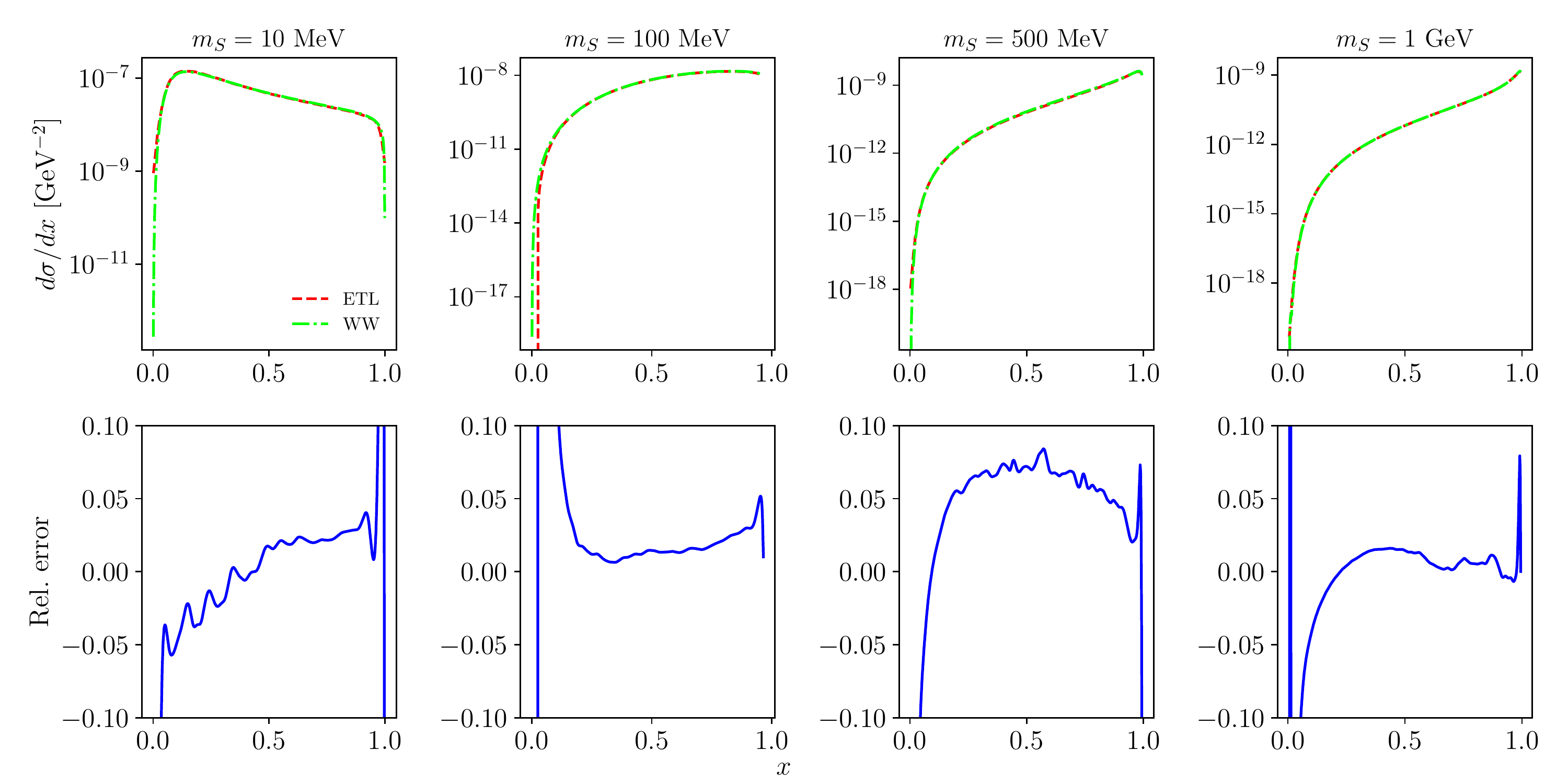}
\caption{\label{fig:comparison_ETL_WW_scalar_ds}
Top: single-differential scalar mediator cross-sections as a function of $x$ in the ETL (red dashed line) and WW approximation (green dashed line) regime for different mass values. The exact approach (ETL) results are obtained numerically through integration by quadrature of the results of Fig. \ref{fig:comparison_ETL_WW_scalar_d2s}. Bottom: relative error, $(\mathcal{O}_{WW}-\mathcal{O}_{ETL})/\mathcal{O}_{ETL}$, between the WW and ETL expressions defined respectively in Eqs. \eqref{eq:etl_d2s} and \eqref{eq:analytical_dsdx_WW}. The mixing strength is $\epsilon_S=10^{-4}$.}
\end{figure}

\begin{figure}[H]
\centering
\includegraphics[width=0.8\textwidth]{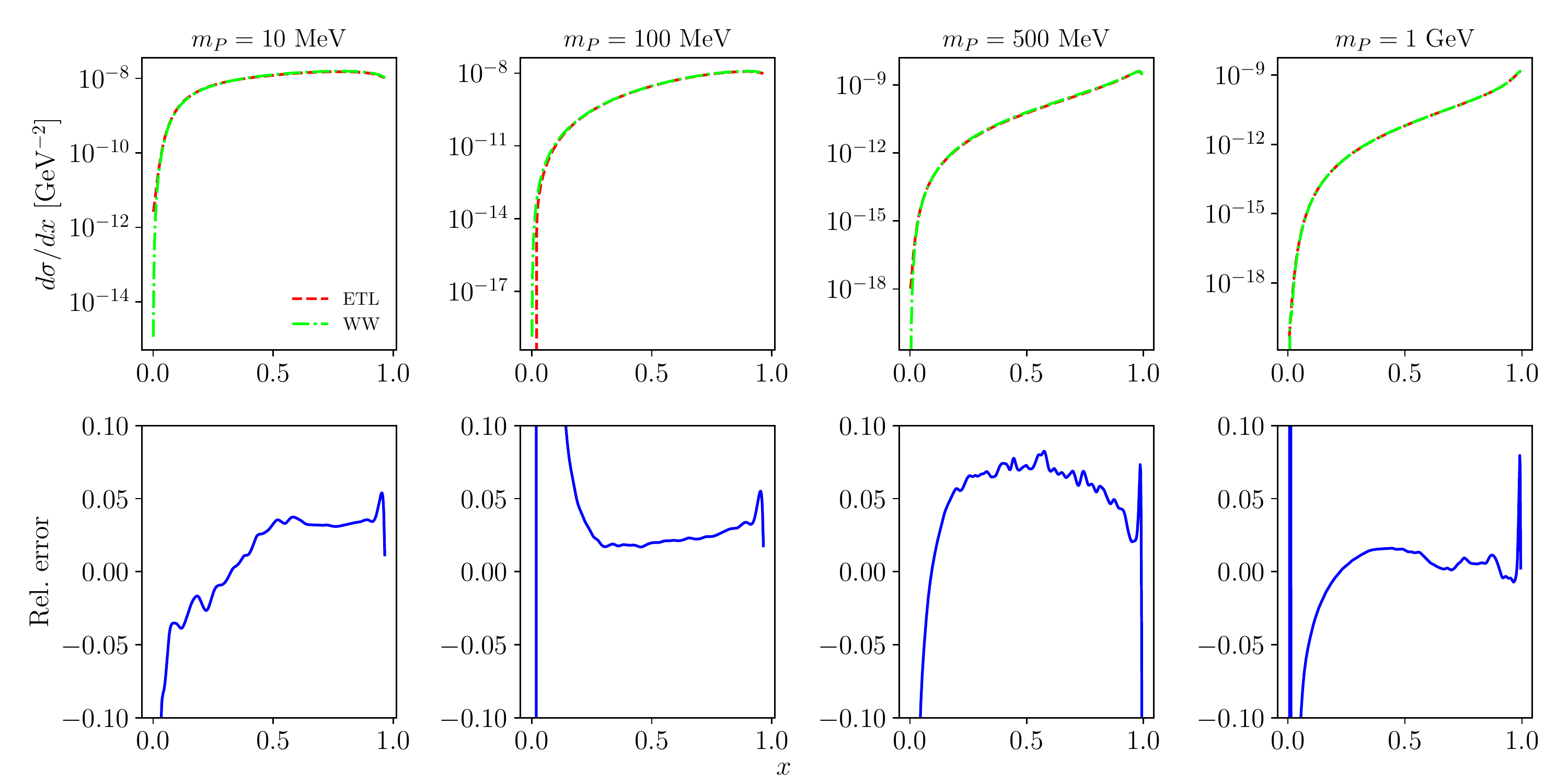}
\caption{\label{fig:comparison_ETL_WW_pseudoscalar_ds}The same as in Fig.~\ref{fig:comparison_ETL_WW_scalar_ds} but for pseudo-scalar mediator $P$.}
\end{figure}
\end{minipage}
\end{widetext}

However, on the boundaries of the fractional 
energy domain,  $x \rightarrow x_\text{min} \ll 1$ and $x\rightarrow x_\text{max}\simeq1$, the relative error can be as 
large as $\gtrsim  \mathcal{O}(5\%)$. After the numerical integration of the differential cross-section, this effect is negligible.  Therefore, the yields of the produced spin-0 bosons can be calculated accurately in WW approach~\cite{Kirpichnikov:2021jev}.

For completeness, the single differential cross-sections with respect to the outgoing muon fractional energy, $y$, and emission angle, $\psi$, are integrated according to Eq. \eqref{eq:WWypsi}. The results are illustrated in Figs. \ref{fig:dsdx_dsdpsi_S} and \ref{fig:dsdx_dsdpsi_P}. Similarly as in the result for the integration over $x$, the integrated differential cross-section relative error is of the order $\lesssim  \mathcal{O}(5\%)$.

\begin{widetext}
\begin{minipage}{\linewidth}
\begin{figure}[H]
    \centering
    \includegraphics[width=0.70\textwidth]{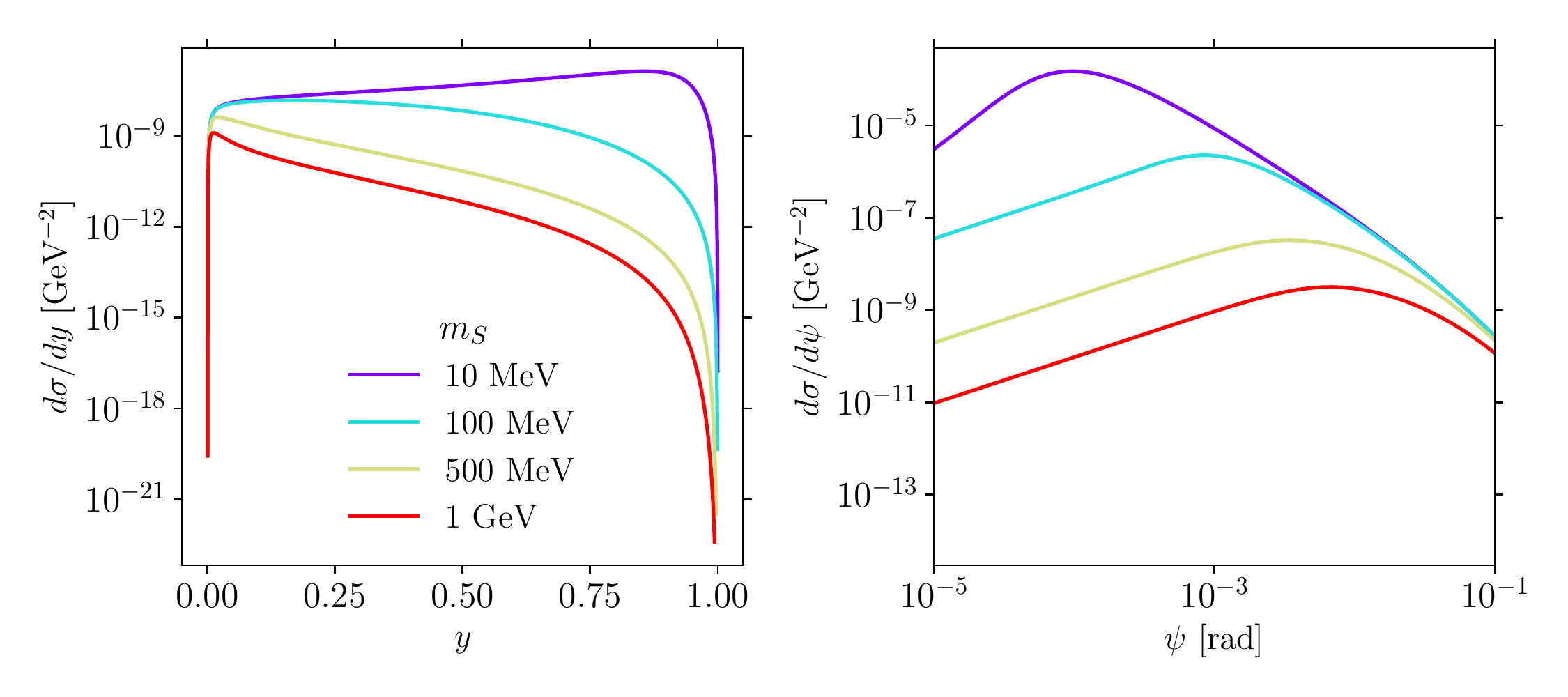}
    \caption{Single-differential scalar mediator cross-sections in WW approximation regime for different mass values (from the top to the bottom line the masses are 10, 100, 500 and 1000 MeV). The mixing strength is $\epsilon_S=10^{-4}$. Left: single-differential cross-section as a function of the outgoing muon fractional energy. Right: single-differential cross-section as a function of the outgoing muon angle.}
    \label{fig:dsdx_dsdpsi_S}
\end{figure}

\begin{figure}[H]
    \centering
    \includegraphics[width=0.70\textwidth]{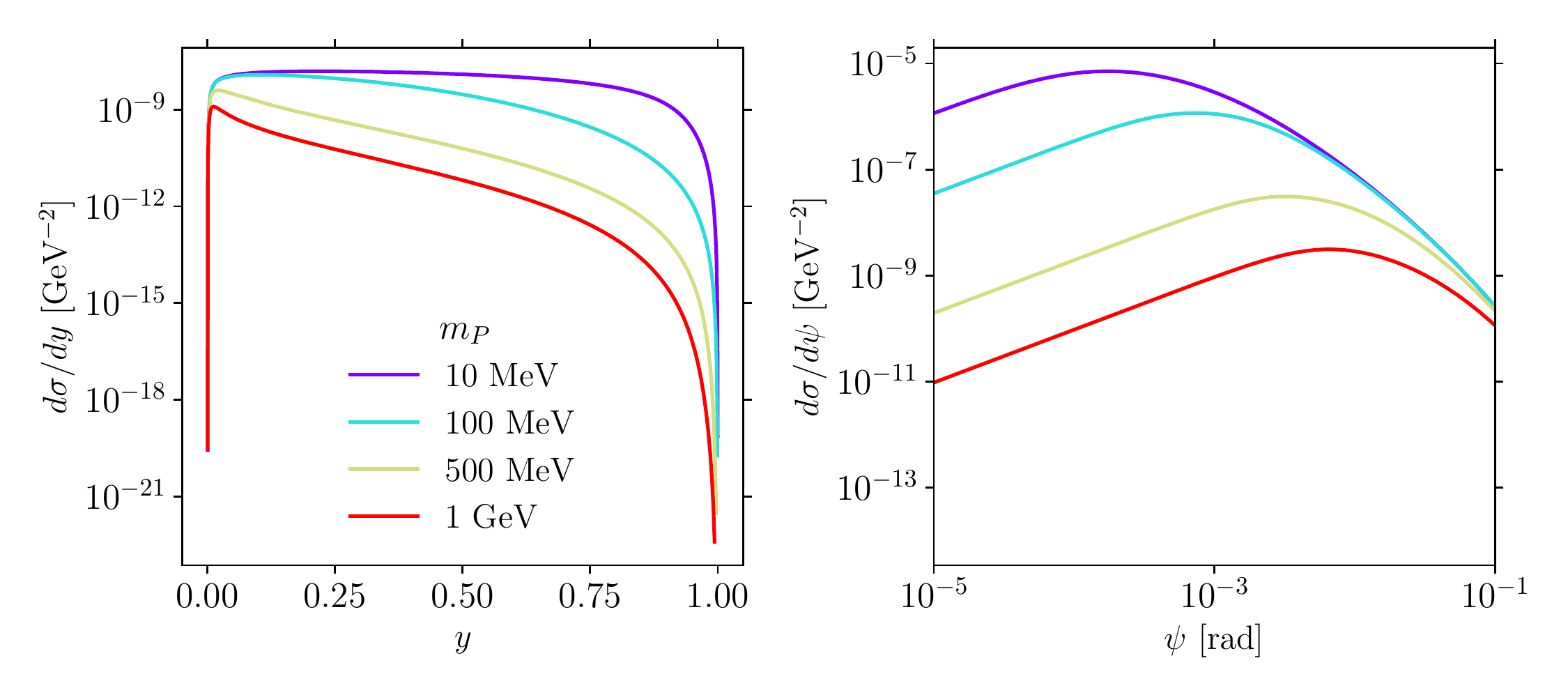}
    \caption{The same as in Fig.~\ref{fig:dsdx_dsdpsi_S} but for pseudo-scalar mediator $P$.}
    \label{fig:dsdx_dsdpsi_P}
\end{figure}
\end{minipage}
\end{widetext}

\section{Projected sensitivities to the mixing strength
\label{BoundsSection}}
The typical estimate of the sensitivity of a given experiment can be computed following the yield formula \cite{Gninenko:2017yus}
\begin{equation}
    N_H=N_\text{MOT}\frac{\rho\mathcal{N}_A}{A}\sum_{i}\sigma_{H}(E_i)\Delta L_i,
\end{equation}
where $\rho$ and $A$ are respectively the target density and its atomic weight, $E_i$ the energy of the muon at the $i-$th step-length $\Delta L_i$ in the target and $N_\text{MOT}$ the number of muons on target. We recall that NA64$\mu$ target is a lead-scintillator sandwich electromagnetic
calorimeter of $40$ radiation lengths $(40 X_0)$~\cite{Sieber:2021fu}. However, for a realistic sensitivity study of one's experiment to New Physics models, including particles propagation through the detectors, the differential cross-sections for the production of light muon-philic mediators are implemented using the interface provided by the fully GEANT4 \cite{GEANT4:2002zbu} compatible DMG4 package. In Fig. \ref{fig:dmg4-yield} the production yields for both cases $H=S$ and $H=P$ as obtained through a realistic GEANT4 simulation of the NA64$\mu$ detectors are shown. The number of signal events $N_H$ are given in the scenario where $\epsilon_H=10^{-4}$. For completeness, the yield for the vector case ($H=V$), for which the similar calculations and implementation were performed previously, is plotted.

\begin{minipage}{\linewidth}    
\begin{figure}[H]
    \centering
    \includegraphics[width=1.0\textwidth]{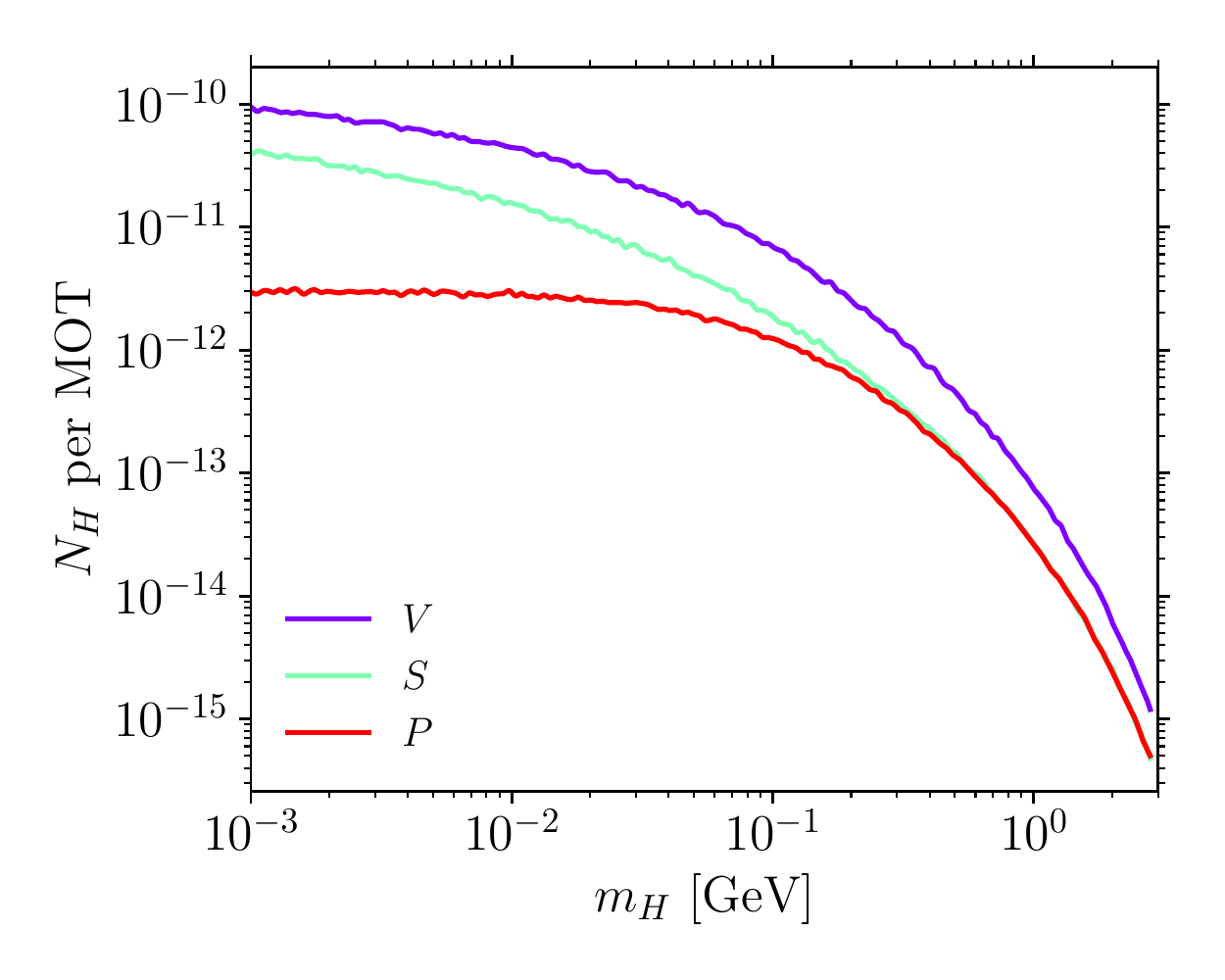}
    \caption{\label{fig:dmg4-yield}Number of light mediators, $N_H$, per muons on target (MOT) as obtained through a full GEANT4 simulation of the NA64$\mu$ target using the DMG4 package ($\epsilon_H=10^{-4}$ is considered in this case). For completeness, in addition to the scalar $(S)$ and pseudo-scalar $(P)$ mediators is also shown the vector $(V)$ case.}
\end{figure}
\end{minipage}

The sensitivity of the experiment in the case where $H=S$ and $H=P$ is shown in Fig. \ref{fig:sensitivity_scalar} 
in the invisible scenario, $S\rightarrow \mbox{invis}.$, in the target parameter space $(m_H,\ g_H)$. Eq. \eqref{eq:analytical_dsdx_WW} is considered, i.e. the WW regime. The limits are calculated at 
90\%~C.~L., requiring $N_{H}\gtrsim  2.3$, and assuming 100\% efficiency and no 
background. In the chosen mass range, both sensitivities for the case $H=S$ and $H=P$ 
yield similar reach as Eqs. \eqref{A2to2S} and \eqref{A2to2P} only differ by the typical factors $m_S^2-4m_\mu^2$ 
and~$m_P^2$.

NA64$\mu$ projected sensitivity for $10^{13}$ MOT can completely probe the New Physics 
contribution to the $(g-2)_\mu$ anomaly for a muon-philic
scalar mediator for masses below 3 GeV. The DM relic predictions have been obtained in \cite{Chen:2018vkr}. The ¨kink¨ present in the DM relic curves arises as at $m_S = 2m_{\mu}$ a new annihilation channel to muons is kinematically accessible. For a similar mass range and MOT, the target parameter space for thermal relic DM with a scalar mediator is fully accessible in the scenario 
where $g_S^\chi=1$ and $m_S=3m_\chi$.  Because of the similar behaviour of the cross-
sections at masses above the muon 
mass ($m_S \gtrsim m_\mu$), the previous statement is also valid in the case of a pseudo-scalar mediator, $H=P$. 
In the near-resonant scenario for which $m_S \simeq  2.1 m_\chi$, 
only the portion of parameter space 
with $m_H\lesssim \mathcal{O}(m_\mu)$ is accessible. Note that for completeness projected sensitivities from both $M^3$ 
phases I and II are shown for $10^{10}$~MOT and $10^{13}$~MOT
respectively given the phase-space values provided in~\cite{Kahn:2018cqs}. Additionally, the ATLAS-HL \cite{Galon:2019owl} and FASER$\nu$(2) \cite{Ariga:2023fjg} expected limits are shown.

\begin{widetext}
\begin{minipage}{\linewidth}    
\begin{figure}[H]
    \centering
    \includegraphics[width=0.45\textwidth]{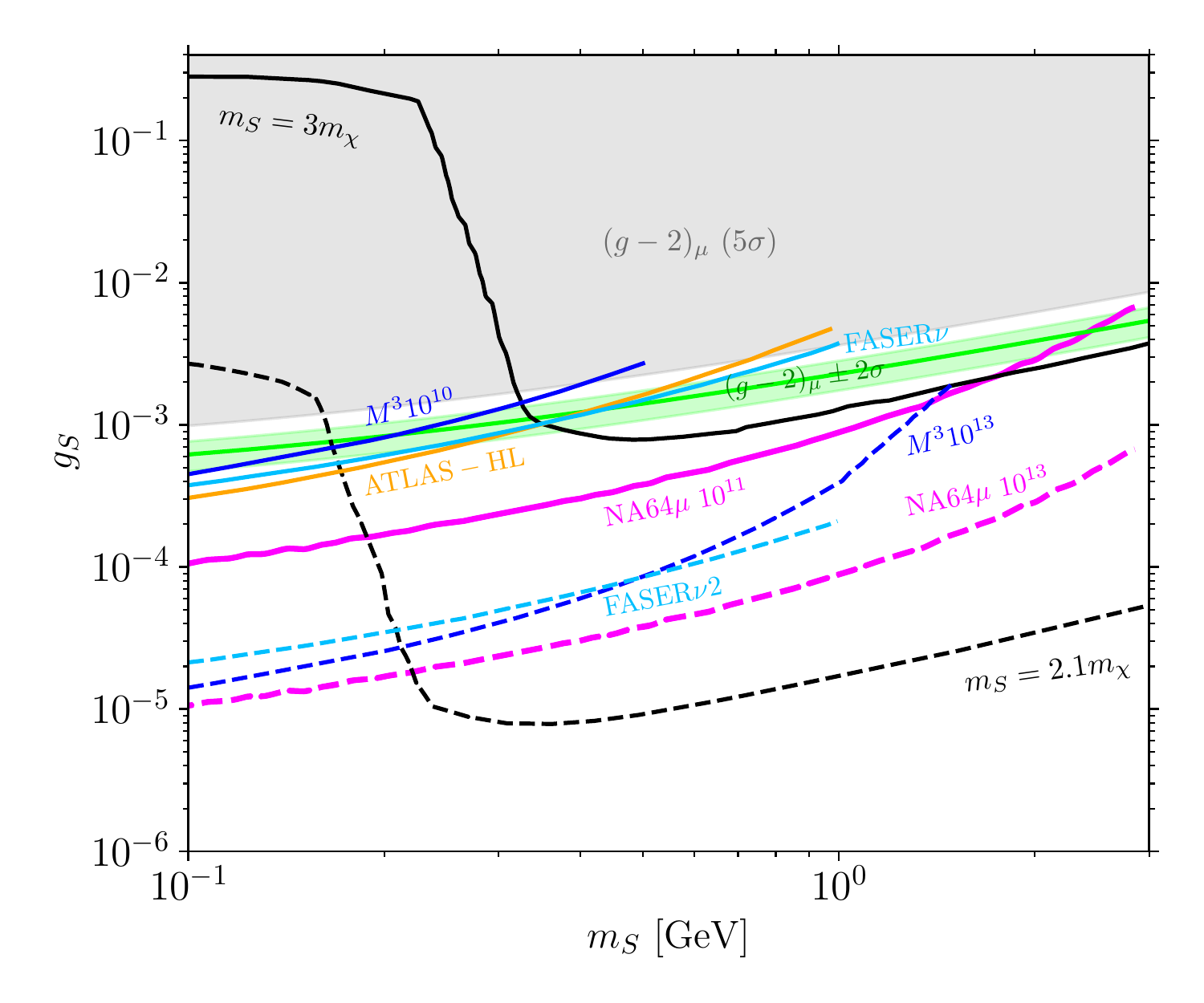}
    \hspace{1mm}
    \includegraphics[width=0.45\textwidth]{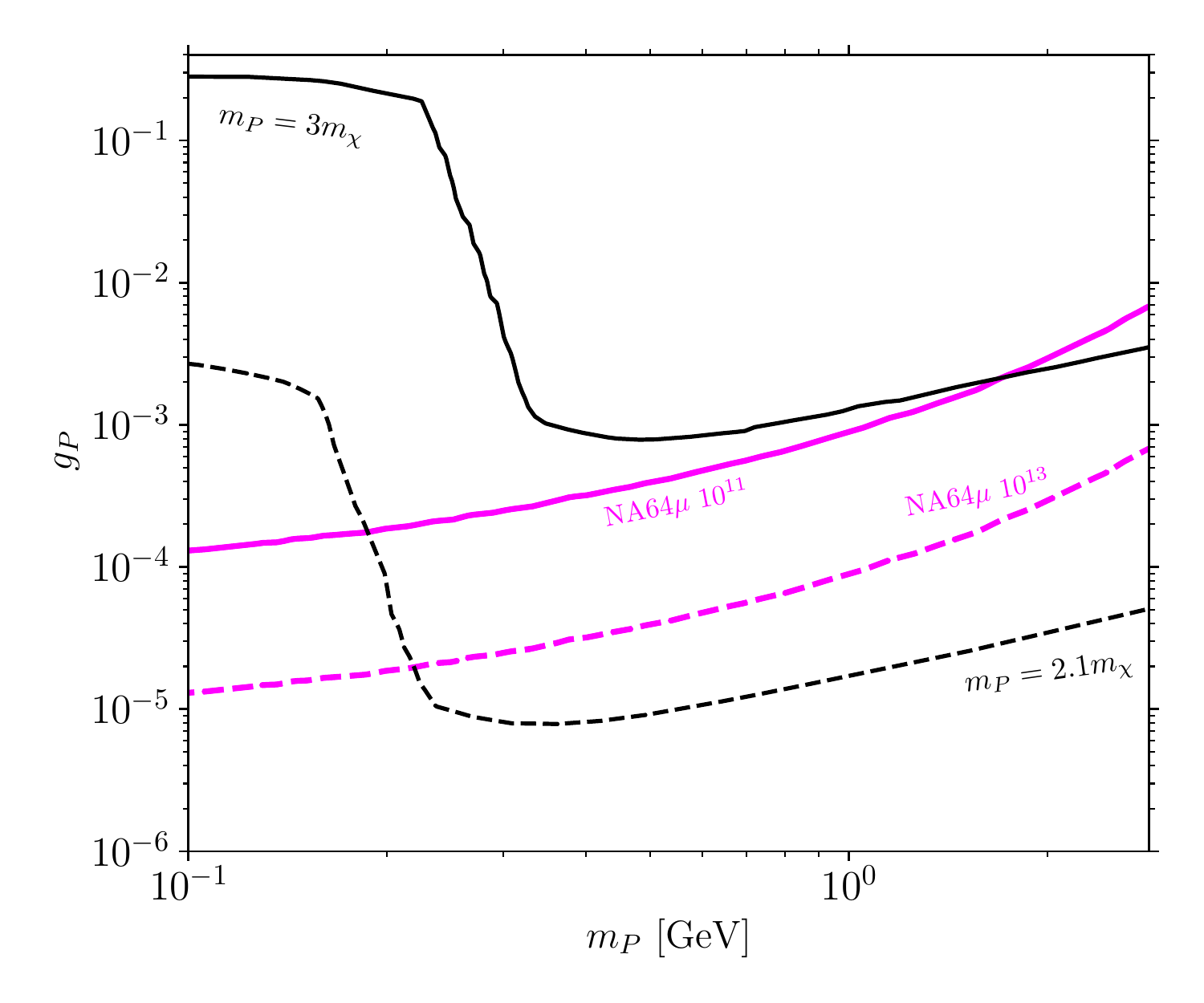}
    \caption{Projected sensitivity in the $(m_H,g_\mu)$ phase space obtained through numerical integration of Eq. \eqref{eq:analytical_dsdx_WW} through \cite{galassi2018scientific} for $10^{11}$ (plain magenta line) and $10^{13}$ (dashed magenta line) MOT. The limits are calculated at 90\% C.L. Left: scalar ($S$) case. Are also shown the $(g-2)_\mu\pm2\sigma$ band following Eq. \eqref{eq:delta_a_S} and the thermal freeze-out target as computed in \cite{Chen:2018vkr} for the scenario $g_\chi=1$ both with $m_S=3m_\chi$ and $m_S\simeq2.1m_\chi$. The $M^3$ experiment phases 1 and 2 (plain and dashed blue lines) are shown for completeness \cite{Kahn:2018cqs} together with the ATLAS HL-LHC analysis at $\mathcal{L}_{\text{LHC}}=3$ ab$^{-1}$ \cite{Galon:2019owl} (orange line) and the FASER$\nu$ projected sensitivity at $\mathcal{L}_{\text{LHC}}=250$ fb$^{-1}$ (light blue line) and $\mathcal{L}_{\text{LHC}}=3\ \text{ab}^{-1}$ (dashed light blue line) \cite{Ariga:2023fjg}. Right: pseudo-scalar ($P$) case.}
    \label{fig:sensitivity_scalar}
\end{figure}
\end{minipage}
\end{widetext}

\section{Summary
\label{SectionSummary}}

In this work, we have derived, based on the work of \cite{Liu:2017htz,Liu:2016mqv,Kirpichnikov:2021jev}, the 
differential cross-sections for spin-0 DM mediator production in fixed target experiments through muon 
bremsstrahlung. We  have shown that the commonly used  Weiszäcker-Williams approximation reproduces well the  
exact-tree-level calculations cross-section with an accuracy at the level of  $\lesssim \mathcal{O}( 5\%)$ in the 
high-energy beam regime.  We have also calculated the $S(P)$ differential 
cross-section as a function of new variables, namely the scattered muon fractional energy and recoil angle, of 
potential importance for Monte Carlo simulations and in the estimate of realistic signal yields in missing momentum experiments.  Additionally, we 
developed an analytical expression of the differential cross section of spin-0 mediators in WW approximation to reduce 
computational time due to numerical  integration. We highlight that the results derived can be relevant for different experiments such as proton beam-dump as NA62, SHADOWS, SHIP and HIKE, muon fixed target experiments such as NA64$\mu$, MUoNE and M3, and future neutrino experiments as DUNE. In this work, we have considered as benchmark the NA64$\mu$ experiment. Finally, our calculations were used to derive the projected sensitivities of the experiment to probe leptophilic scenarios.  Our results demonstrate the potential of muon fixed target experiments to explore
a broad coupling and mass  region parameter space of spin-0 DM mediators, including the DM relic and the $(g-2)_\mu$  anomaly favoured parameter space.

\section*{Acknowledgments}
We acknowledge the members of the NA64 collaboration for fruitful discussions.
DK and IV are indebted to R.~Dusaev, D.~Forbes, Y.~Kahn, V.~Lyubovitskij,   A.~Pukhov, and A.~Zhevlakov
for helpful  suggestions and correspondence.  The work of D. V. Kirpichnikov on calculation of spin-0 mediator emission 
is  supported  by the Russian Science Foundation RSF grant 21-12-00379. The work of P. Crivelli and H. Sieber is 
supported by SNSF and ETH Zurich Grants No. 186181, No. 186158 and No. 197346 (Switzerland). The work of L. Molina-Bueno is supported by SNSF Grant No. 186158 (Switzerland), RyC-030551-I, and PID2021-123955NA-100 funded by MCIN/AEI/ 10.13039/501100011033/FEDER, UE (Spain).



\appendix
\section{Special functions
\label{AppSect}}
In this section, we collect the coefficients needed for the analytical differential cross-section calculation 
	\begin{multline*}
		C^{\chi}_{1}  
	=
		\frac{Z^2 t_d^2}{(t_a - t_d)^3}
		\left(
		 	\frac{t_d(t_a - t_d)}{t_d + t_\text{max}}
		+	\frac{t_a(t_a - t_d)}{t_a + t_\text{max}}
		- \right. \\ \left. -	
		    2 (t_a - t_d)
		+	(t_a + t_d) 
		\ln{\left( \frac{t_d + t_\text{max}}{t_a + t_\text{max}}\right) }
		\right),
	\end{multline*}
	\[
		C^{\chi}_{2}  
\!	=\!
		\frac{Z^2 t_d^2 g^2 }{(t_a\! - \!t_d)^3} \!\!
		\left(\!
		\frac{t_a \!- \!t_d}{t_d \! + \! t_\text{max}}
		\!+	\! \frac{t_a \!- \! t_d}{t_a \! +\! t_\text{max}}
\!		+\!	2 \ln{\!\left( \!\frac{t_d \!+ \!t_\text{max}}{t_a\! + \!t_\text{max}}\!\right) }
		\!\! \right), 
	\]
	\[
		C^{\chi}_{3}  
	=
	-	\frac{Z^2 t_d^2 (t_a + t_d)	}{(t_a - t_d)^3},
	\quad
		C^{\chi}_{4}  
	=
	-	\frac{2 Z^2 t_d^2 g^2 }{(t_a - t_d)^3}.
	\]
The coefficients for the resulting cross section (\ref{eq:analytical_dsdx_WW}) are 
	\begin{multline}
		I^H_1(x,u)
	=
		C^{H}_{1} C^{\chi}_{2}  u
	+	C^{H}_{2} C^{\chi}_{2}  \frac {\ln{(u^2)} }{2}
	- \\ -
		\frac{C^{H}_{1} C^{\chi}_{1}  
			 + C^{H}_{3} C^{\chi}_{2} }{u}
	-	\frac{C^{H}_{2} C^{\chi}_{1} }{2u^2}
	-	\frac{C^{H}_{3} C^{\chi}_{1} }{3u^3},
	\end{multline}
	\begin{multline}
		I^H_2(x,u)
	=
		\frac{C^{H}_{3} C^{\chi}_{3} }{3}
	\cdot \\ \cdot
		\left\lbrace  
		-	\frac{1}{u^3}
			f_1(x, u)
		-	2 \frac{g^2 }{u} 
			\left(\frac{1}{t_d} - \frac{1}{t_a} \right) 
		-	f_{2}(x, u, 4)
		\right\rbrace, 
	\end{multline}
	\begin{equation}
		I^H_3(x,u)
	=
		\frac{C^{H}_{2} C^{\chi}_{3} }{2}
		\left\lbrace  
			f_3(x, u) 
		-	\frac{1}{u^2}
			f_1(x, u) 
		\right\rbrace, 
	\end{equation}
	\begin{equation}
		I^H_4(x,\!u)
	\!=\!
		\left( 	
		    C^{H}_{1} \! C^{\chi}_{3}  
		\!+ \! 	C^{H}_{3} \! C^{\chi}_{4} 
		\right) \!\!
		\left\lbrace 
		 	f_{2}(x,\! u,\! 2) 	\!-\!	\frac{1}{u}
			f_1(x, \! u)\!
		\right\rbrace, 
	\end{equation}
	\begin{multline}
		I^H_5(x,u)
	=
		\frac{C^{H}_{2} C^{\chi}_{4} }{2}
	\cdot \\ \cdot
		\left\lbrace
			\! \ln{ \!\left( \! \frac{t_d}{t_a} \! \right) }
\!			\ln{(u^2)}
		\! - \!	Li_2\left(\! - \frac{g^2  u^2}{t_d} \!\right)
	\!	+ \!	Li_2\left( \!- \frac{g^2  u^2}{t_a}\! \right) 	
	\! \right\rbrace, 
	\end{multline}
	\begin{equation}
		I^H_6(x,u)
	=
		C^{H}_{1} C^{\chi}_{4} 
		\left\lbrace
			u f_1(x, u) 
		+	f_{2}(x, u, 0)
		\right\rbrace, 
	\end{equation}
	where $Li_2(x)$ is a polylogarithm  and  the auxiliary  functions are
	\begin{equation}
		f_1(x, u)
	=
		\ln{\left( \frac{u^2 + b}{u^2 + a}\right) },
	\quad
	    b = \frac{t_d}{g^2 },
	\quad
	    a = \frac{t_a}{g^2 }, 
	\end{equation}
	\begin{align}
	&	f_2(x, u, n)
	=
		2 \left( \frac{1}{b}\right)^{\frac{n-1}{2}}
		\arctan{\left( \frac{u}{\sqrt{b}} \right) } \nonumber
  \\
&	-	2 \left( \frac{1}{a}\right)^{\frac{n-1}{2}}
		\arctan{\left( \frac{u}{\sqrt{a}} \right) },
	\end{align}
	\begin{align}
&		f_3(x, u)
	=   
		 \frac{1}{b} 
		\ln{\left( \frac{u^2 / g^2  }{u^2 + b}\right) }  -	 \frac{1}{a}
		\ln{\left( \frac{u^2 / g^2 }{u^2 + a}\right) }.
	\end{align}

\newpage

\newpage
\bibliography{bibl}	

\end{document}